\documentclass[submission,copyright]{eptcs}
\providecommand{\event}{ICE 2011} 

\newcommand{\ms}[1]{|{#1}|^{2}}

\newcommand{\ket}[1]{|#1\rangle}
\newcommand{\nil}{\mathbf{0}}
\renewcommand{\parallel}{\mathbin{\mid}}
\newcommand{\inp}[2]{{#1}?{[#2]}}
\newcommand{\outp}[2]{{#1}!{[#2]}}

\newcommand{\new}{\mathsf{new}\ }
\newcommand{\qbit}{\mathsf{qbit}\ }

\newcommand{\qgate}[1]{\mathsf{#1}}
\newcommand{\pname}[1]{\mathit{#1}}
\newcommand{\action}[1]{\{#1\}}
\newcommand{\trans}[2]{{#1}\mathbin{*\!\!=}{#2}}
\newcommand{\sep}{\,.\,}

\newcommand{\measure}{\mathsf{measure}\ }

\title{Formal Analysis of Quantum Systems using Process Calculus\thanks{Partially supported by the UK EPSRC:
    \emph{Network on Semantics of Quantum Computation} (EP/E00623X/1)
    and \emph{Quantum Computation: Foundations, Security, Cryptography
      and Group Theory} (EP/F020813/1). and the EU Sixth Framework
  Programme (Project \emph{SecoQC: Development of a Global Network for
  Secure Communication based on Quantum Cryptography}).} }
\author{Timothy A. S. Davidson
\institute{Department of Computer Science\\
University of Warwick, UK}
\email{T.Davidson@warwick.ac.uk}
\and
Simon J. Gay
\institute{School of Computing Science\\
University of Glasgow, UK}
\email{Simon.Gay@glasgow.ac.uk}
\and 
Rajagopal Nagarajan
\institute{Department of Computer Science\\
University of Warwick, UK}
\email{R.Nagarajan@warwick.ac.uk}
} 
\def\titlerunning{Formal Analysis of Quantum Systems using Process Calculus}
\def\authorrunning{T. A. S. Davidson, S. J. Gay and R. Nagarajan}
\begin{document}
\maketitle

\begin{abstract}
  Quantum communication and cryptographic protocols are well on the
  way to becoming an important practical technology. Although a large
  amount of successful research has been done on proving their
  correctness, most of this work does not make use of familiar
  techniques from formal methods such as formal logics for specification,
  formal modelling languages, separation of levels of abstraction,
  and compositional analysis.  We argue that these
  techniques will be necessary for the analysis of large-scale systems
  that combine quantum and classical components, and summarize the
  results of initial investigation using behavioural
  equivalence in process calculus. This paper is a summary of Simon
  Gay's invited talk at ICE'11.
\end{abstract}

\section{Introduction}
Quantum computing and quantum communication (more generally,
\emph{quantum information processing}) appear in the media from time
to time, usually with misleading statements about the principles of
quantum mechanics, the nature of quantum information processing, and
the power of quantum algorithms. In this article, we begin by
clarifying the fundamental concepts of quantum information and
discussing what quantum computing systems are and are not capable
of. We then outline several reasons for being interested in quantum
information processing. Moving on to the main theme, we motivate the
application of formal methods, including process calculus and
model-checking, to quantum systems. Finally, we focus on a particular
quantum process calculus called Communicating Quantum Processes (CQP),
illustrate it by defining a quantum teleportation protocol, and
describe recent results about behavioural equivalence.

\section{What is quantum information processing?}
The idea of quantum information processing (QIP) is to represent
information by means of physical systems whose behaviour \emph{must}
be described by the laws of quantum physics. Typically this means very
small systems, such as a single atom (in which the spin state, up or
down, gives the basic binary distinction necessary for digital
information representation) or a single photon (in which polarization
directions are used). Information is then processed by means of
operations that arise from quantum physics. Quantum mechanics leads to
several fundamental properties of quantum information, which between
them lead to various counter-intuitive effects and the possiblity of
behaviour that cannot occur in classical systems.

\subsection{Superposition}
The state of a classical bit is either $0$ or $1$. The state of a
quantum bit (qubit) is $\alpha\ket{0}+\beta\ket{1}$, where $\ket{0}$
and $\ket{1}$ are the \emph{basis states}. In general, $\alpha$ and
$\beta$ are complex numbers, and if both of them are non-zero
then the state is a \emph{superposition}, for example
$\frac{1}{\sqrt{2}}\ket{0}-\frac{1}{\sqrt{2}}\ket{1}$. It is not
correct to say, as often stated in the media, that a qubit can be in
two states at once. It is in one state, but that state may be a
superposition of the basis states.

\subsection{Measurement}
It is not possible to inspect the contents of a quantum state. The
most we can do is a measurement. Measuring a qubit that is in state
$\alpha\ket{0}+\beta\ket{1}$ has a random result: with probability
$\frac{1}{\ms{\alpha}}$ the result is $0$, and with probability
$\frac{1}{\ms{\beta}}$ the result is $1$. After the measurement, the
qubit is in the basis state corresponding to the result.

\subsection{Operations on a superposition}
An operation acts on every basis state in a superposition. For
example, starting with the three-qubit state
$\frac{1}{2}\ket{000}+\frac{1}{2}\ket{010}-\frac{1}{2}\ket{110}-\frac{1}{2}\ket{111}$
and applying the operation ``invert the second bit'' produces the
state
$\frac{1}{2}\ket{010}+\frac{1}{2}\ket{000}-\frac{1}{2}\ket{100}-\frac{1}{2}\ket{101}$. This
is sometimes known as \emph{quantum parallelism} and in the media it
is often described as carrying out an operation simultaneously on a
large number of values. However, it is not possible to discover the
results of these simultaneous operations. A measurement would produce
just one of the basis states. This is absolutely not a straightforward
route to ``parallelism for free''.

\subsection{No cloning}
It is not possible to define an operation that reliably makes a
perfect copy of an unknown quantum state. This is known as the
\emph{no cloning theorem}. It contrasts sharply with the classical
situation, where the existence of uniform copying procedures is one of
the main advantages of digital information. Every word in the
statement of the no cloning theorem is significant. For example, with
the knowledge that a given qubit is either $\ket{0}$ or $\ket{1}$, it
is possible to discover its state (by means of a simple measurement)
and then set another qubit to the same state, thus creating a copy. It
is also possible in general to create approximate copies, or to copy
with a certain probability of perfect success but a certain
probability of complete failure. It is possible to transfer an unknown
quantum state from one physical carrier to another, but the process
destroys the original state. This is known as \emph{quantum
  teleportation}, and we will return to it later.

\subsection{Entanglement}
The states of two or more qubits can be correlated in a way that is
stronger than any possible classical correlation. An example is the
two-qubit state
$\frac{1}{\sqrt{2}}\ket{00}+\frac{1}{\sqrt{2}}\ket{11}$. Measuring
either qubit produces, with equal probability, the state $\ket{00}$ or
$\ket{11}$. Measuring the other qubit is then guaranteed to produce the
  same result as the first measurement. This correlation is preserved
  by quantum operations on the state, in a way that cannot be
  reproduced classically. This phenomenon is called
  \emph{entanglement} and it is a key resource for quantum algorithms
  and communication protocols.

\section{Quantum algorithms and protocols}
We will now summarize a few algorithms and protocols in which quantum
information processing has a clear advantage over classical
information processing. This list is not complete; in particular,
there are many more cryptographic protocols than we mention
here. Teleportation is not included here as we will discuss it in more
detail later.

\subsection{The Deutsch-Jozsa algorithm}
Suppose an unknown function $f:\{0,1\}^n\rightarrow\{0,1\}$, is given
as a black box, together with information that $f$ is either
\emph{constant} or \emph{balanced} (meaning that its value is $0$ for
exactly half of its inputs). The Deutsch-Jozsa algorithm
\cite{DeutschD:rapspq} works out whether $f$ is constant or balanced,
with only one evaluation of $f$. Classically, $2^{n-1}+1$ evaluations
would be required in the worst case.

\subsection{Shor's algorithm}
Shor's algorithm \cite{ShorPW:algqcd} is for integer
factorization. Its complexity is $O((\log n)^3)$, whereas the best
known classical algorithm has complexity $O(e^{(\log
  n)^\frac{1}{3}(\log\log n)^\frac{2}{3}})$. The RSA cryptosystem
relies on the unproven assumption that factorization is intractable,
so a practical implementation of Shor's algorithm would threaten
current information security technology. Note, however, that there is
no proved non-polynomial lower bound for classical factorization
algorithms, and factorization is not believed to be an NP-complete
problem. Media reports about quantum computing often give the
impression that quantum computers can solve NP-complete problems
efficiently, but there is no evidence for this statement.

\subsection{Grover's algorithm}
Grover's algorithm \cite{GroverL:fasqma} finds an item in an
unstructured list of length $n$, taking time
$O(\sqrt{n})$. Classically, every item must be inspected, requiring
$O(n)$ time on average.

\subsection{Quantum key distribution}
Quantum key distribution (QKD) protocols, such as the BB84
\cite{BennettCH:quacpd} protocol of
Bennett and Brassard, generate shared cryptographic keys which can
then be used with a classical encryption technique such as a one-time
pad. QKD is secure against any attack allowed by the laws of quantum
mechanics, including any future developments in quantum
computing. Essentially, secrecy of the key is guaranteed by the no
cloning theorem: an attacker cannot make a perfect copy of any
information that she intercepts while the protocol is running, and
therefore either receives negligible information or reveals her presence.

\section{Why is QIP interesting, and will it become practically
  significant?}
There are several reasons to be interested in quantum information
processing. First, the subject is really about understanding the
information-processing power permitted by the laws of physics, and
this is a fundamental scientific question. Second, quantum algorithms
might help to solve certain classes of problem more
efficiently; if, however, NP-complete problems cannot be solved
efficiently even by a quantum computer, then understanding why not is
also a question of fundamental interest. Third, quantum cryptography
provides a neat answer, in advance, to any threat that quantum
computing might pose to classical cryptography. Fourth, as integrated
circuit components become smaller, quantum effects become more
difficult to avoid. Quantum computing might be necessary in order to
continue the historical trend of miniaturization, even if it offers no
complexity-theoretic improvement. Finally, Feynman suggested that
quantum computers could be used to simulate complex (quantum) physical
systems whose behaviour is hard to analyze classically. 

Will QIP become practically significant? Some aspects are already
practical: there are companies selling QKD systems today. Whether or
not there is a real demand for quantum cryptography remains to be
seen, but it seems likely that the promise of absolute security will
attract organizations that feel they cannot take any chances. Quantum
computing seems to be feasible in principle, although there are still
formidable scientific and engineering challenges. But many
experimental groups are working hard, and physicists and engineers are
very clever. Remember that in 1949 the statement ``In the future,
computers may weigh no more than 1.5 tonnes'' was a very speculative
prediction.

\section{Formal methods for QIP}
There is no doubt about the correctness of quantum algorithms and
protocols. Simple protocols such as teleportation can be checked with
a few lines of algebra, Shor's and Grover's algorithms have been
extensively studied, and Mayers \cite{MayersD:uncsqc} and others have
proved the security of quantum key distribution. But what about
\emph{systems}, which are constructed from separate components and
combine quantum and classical computation and communication?
Experience in classical computing science has shown that correctness
of a complete implemented system is a very different question from
correctness of the idealized mathematical protocol that it claims to
implement. This is the \emph{raison d'\^etre} of the field of formal
methods.

Nagarajan and Gay \cite{NagarajanR:forvqp} suggested applying formal
methods to quantum systems, with the same motivation as for classical
systems:
\begin{itemize}
\item \emph{formal modelling languages}, for unambiguous definitions;
\item analysis of \emph{systems}, rather than idealized situations;
\item \emph{systematic verification methodologies}, rather than
  \emph{ad hoc} reasoning;
\item the possibility of \emph{tool support}.
\end{itemize}
We have been working on two strands: quantum process calculus
\cite{GaySJ:comqp,GaySJ:typtcq}, most recently in collaboration with
Davidson \cite{DavidsonT:forvtq}, and model-checking, in collaboration
with Papanikolaou
\cite{GaySJ:qmcmcq,GaySJ:spevqp,PapanikolaouNK:modcqp}. In general
these approaches are not mutually exclusive. However, our work on
process calculus has focussed on the development of basic theory,
leading up to the definition of behavioural equivalence; our work on
model-checking uses a different style of specification language, more
closely related to Promela. Some recent work \cite{DavidsonT:modccq}
makes connections between the two themes.

Other approaches to quantum process calculus include Jorrand and
Lalire's QPAlg \cite{JorrandP:towqpa} and Ying \emph{et al.}'s qCCS \cite{YingM:algqp}.

\section{Quantum teleportation in CQP}
Teleportation \cite{BennettCH:teluqs} is a protocol for transferring
an unknown qubit state from one participant, Alice, to another,
Bob. The protocol uses classical communication --- in fact,
communication of just two classical bits --- to achieve the transfer
of a quantum state which is specified by two complex numbers. The
trick is that there must be some pre-existing entanglement, shared by
Alice and Bob.

Let $x$ and $y$ refer to two qubits that, together, are in the
entangled state
$\frac{1}{\sqrt{2}}\ket{00}+\frac{1}{\sqrt{2}}\ket{11}$. Let $u$ be a
qubit in an unknown state, that is given to Alice. The protocol
consists of the following steps.
\begin{enumerate}
\item Alice applies the \emph{controlled not} operator to $u$ and
  $x$. This is a two-qubit operator whose effect on each basis state
  is to invert the second bit if and only if the first bit is $1$.
\item Alice applies the \emph{Hadamard} operator to $x$. This operator
  is a change of basis from $\{\ket{0},\ket{1}\}$ to
  $\{\frac{1}{\sqrt{2}}(\ket{0}+\ket{1}),
    \frac{1}{\sqrt{2}}(\ket{0}-\ket{1})$.
\item Alice measures $u$ and $x$, obtaining a two-bit classical
  result.
\item Alice sends this two-bit classical value to Bob.
\item Bob uses this classical value to determine which of four
  operators should be applied to $y$.
\item The state of $y$ is now the original state of $u$ (and $u$ has
  lost its original state and is in a basis state).
\end{enumerate}
Although the measurement in step 3 has a probabilistic result, the use
of the classical value to determine a compensating operation in step 5
means that the complete protocol is deterministic in its effect on the
state of Bob's qubit.

The following definitions in the process calculus CQP (Communicating
Quantum Processes) \cite{GaySJ:comqp,GaySJ:typtcq} model the teleportation
protocol. $\pname{Alice}$, $\pname{Bob}$ and $\pname{Teleport}$ are
processes; $q$ is a formal parameter representing a qubit;
$\mathit{in}$, $\mathit{out}$, $a$ and $b$ are formal parameters
representing channels; $c$ is a private channel; $x$, $y$ are local
names for freshly allocated qubits, which will be instantiated with
the names of actual qubits during execution. The language is based on
pi-calculus and most of the syntax should be familiar.
\[
\begin{array}{l}
\pname{Alice}(q,\mathit{in},\mathit{out}) 
= 
\inp{\mathit{in}}{u}\sep\action{\trans{u,q}{\qgate{CNot}}}\sep\action{\trans{u}{\qgate{H}}}\sep\outp{\mathit{out}}{\measure
u,q}\sep\nil \\
\pname{Bob}(q,\mathit{in},\mathit{out}) 
=  \inp{\mathit{in}}{r}\sep\action{\trans{y}{\sigma_{r}}}\sep\outp{\mathit{out}}{y}\sep\nil
\\
\pname{Teleport}(a,b) 
= 
(\qbit
x,y)(\action{\trans{x}{\qgate{H}}}\sep\action{\trans{x,y}{\qgate{CNot}}}\sep(\new c)(\pname{Alice}(x,a,c) \parallel \pname{Bob}(y,c,b))
\end{array}
\]
In $\pname{Teleport}$, the actions before $(\new c)$ put the qubits
$x$ and $y$ into the necessary entangled state. In order to help with
writing a specification, $\pname{Alice}$ is given the qubit to be
teleported as a message on channel $\mathit{in}$, and at the end of
the protocol, $\pname{Bob}$ outputs the final qubit on $\mathit{out}$.

CQP has an operational semantics defined by labelled transition rules;
it also has a type system in which the no cloning theorem is
represented by linear typing. The example above, for simplicity, does
not include type declarations.

The desired behaviour of teleportation is that a qubit (quantum state)
is received on $a$ and the same quantum state is sent on $b$; the
protocol should behave like an identity operation:
\[
\pname{Identity}(c,d) = \inp{c}{x}\sep\outp{d}{x}\sep\nil
\]
We can now write a 
 specification of teleportation:
\[
\pname{Teleport}(c,d) \cong \pname{Identity}(c,d)
\]
where $\cong$ is a behavioural equivalence. 
Equivalent
processes cannot be distinguished by any observer: they output the
same values in the same circumstances, they produce the same
probability distributions of measurement results, and in general
interact in the same way with their environment.

As usual, we would like behavioural equivalence
to be a congruence:
\[
\forall P,Q,C.\quad P\cong Q \Rightarrow C[P] \cong C[Q]
\]
where $C$ is a process context. Congruence supports equational
reasoning, and the universal composability properties defined by
Canetti \cite{CanettiR:unicsn} in a different setting.
Developing a congruence for a quantum process calculus was an open
problem for several years \cite{LalireM:relqpb}, but very recently we
have defined a congruence for CQP \cite{DavidsonT:forvtq} and Feng
\emph{et al.} have independently defined one for qCCS
\cite{FengY:bisqp}. Our equivalence is a form of probabilistic
branching bisimulation \cite{TrckaN:brabcp}, with appropriate extensions to
deal with the quantum state. We have proved that the specification of
teleportation is satisfied.

\section{Conclusion}
We have outlined the principles of quantum information processing, and
argued that formal methods will be necessary in order to guarantee the
correctness of practical quantum systems. We have illustrated a
particular approach --- specification and verification via behavioural
equivalence in quantum process calculus --- with reference to quantum
teleportation.

Future work on the theoretical side will include the development of
equational axiomatizations of behavioural equivalence in CQP, and the
automation of equivalence checking. On the practical side, we intend
to work on more substantial examples including cryptographic systems.

\bibliographystyle{eptcs}
\bibliography{main}
\end{document}